\documentclass[reprint,twocolumn,jcp,longbibliography,superscriptaddress]{revtex4-1}
\usepackage{graphics,graphicx}
\usepackage[T1]{fontenc}

\begin{document}

\title{Nonrigidity effects --- a missing puzzle piece in the description of low-energy anisotropic molecular collisions}

\author{Mariusz Pawlak}
\email{e-mail: teomar@chem.umk.pl}
\affiliation{Faculty of Chemistry, Nicolaus Copernicus University in Toru\'n, Gagarina~7, 87-100~Toru\'n, Poland}
\author{Piotr S. {\.Z}uchowski}
\email{e-mail: pzuch@fizyka.umk.pl}
\affiliation{Faculty of Physics, Astronomy and Informatics, Nicolaus Copernicus University in Toru\'n, Grudzi\k{a}dzka~5, 87-100~Toru\'n, Poland}
\author{Nimrod Moiseyev}
\affiliation{Schulich Faculty of Chemistry and Department of Physics, Technion--Israel Institute of Technology, Haifa 32000, Israel}
\author{Piotr Jankowski}
\affiliation{Faculty of Chemistry, Nicolaus Copernicus University in Toru\'n, Gagarina~7, 87-100~Toru\'n, Poland}

\begin{abstract}
\noindent
Cold collisions serve as a very sensitive probe of the interaction potential. In the recent study of Klein {\em et al.} ({\it Nature Phys.} 13, 35-38 (2017)) the one-parameter scaling of the interaction potential was necessary to obtain agreement between theoretical and observed patterns of the orbiting resonances  for excited metastable helium atoms colliding with hydrogen molecules. Here we show that the effect of nonrigidity of the H$_2$ molecule on the resonant structure, absent in the previous study, is critical to predict correct positions of the resonances in that case. We have complemented the theoretical description of the interaction potential and revised  reaction rate coefficients by proper inclusion of the flexibility of the molecule. The calculated reaction rate coefficients are in remarkable agreement with the experimental data without empirical adjustment of the interaction potential. We have shown that even state-of-the-art calculations of the interaction energy cannot ensure agreement with the experiment if such an important physical effect as flexibility of the interacting molecule is neglected. Our findings about the significance of the nonrigidity effects can be especially crucial in cold chemistry, where the quantum nature of molecules is pronounced.
\end{abstract}

\maketitle

The breakthrough in controlling the movement and internal degrees of molecules with external fields~\cite{vdMeerakker_2008}, which started about 20 years ago, currently allows to  study the collisions, reactivity and properties of molecules in very cold regime. Such molecules provide new and propitious prospects in precision spectroscopy, fundamental physics, astrochemistry, and quantum engineering. Understanding of quantum effects, resonance phenomena, and reaction dynamics in a low-energy range opens the gate to design and create materials with unusual functionality and elements of quantum computers~\cite{Cote_2014, Bohn_Science_2017}. Cold collision experiments realized by merging two supersonic beams have become an important technique for studying chemical reactions in temperatures near 1~K~\cite{Henson_Ed_Science_2012, Ed_NatChem_2014, Jankunas_JCP_2014, Jankunas_JCP_2015, Ed_NatChem_2015, Klein:16} in which unseen earlier quantum features, such as resonances or interference, are revealed. The group of Narevicius performed the first experiment in that field focused mainly on the Penning ionization (PI) process of colliding hydrogen isotopologues with excited metastable helium atoms~\cite{Henson_Ed_Science_2012, Ed_NatChem_2014, Ed_NatChem_2015, Klein:16}. In such a reaction, an electron is moved from the molecule to the only partially-occupied orbital of excited atom, then the initially excited electron of the atom is kicked out of the system. Three products occur: the atom in the ground state, the molecular ion in the ground state, and a free electron. These colliding systems are of great interest to astrophysicists and astrochemists studying conditions and reactions in outer space. Hydrogen and helium are the most abundant elements, whereas molecular hydrogen is the most common molecular species in the Universe~\cite{Dalgarno_2000, Ed_NatChem_2015}. The first observation of metastable helium atom in the atmosphere of one of exoplanets~\cite{Spake_2018} has boosted its importance for astrochemistry, and one can expect that its interaction with the omnipresent hydrogen molecule will be carefully investigated.

Recently, Narevicius and co-workers directly probed the anisotropy in atom--molecule interactions through orbiting resonances by changing the rotational state of the molecule~\cite{Klein:16}. That work reveals a crucial role of the anisotropy of the interaction energy, due to various orientations of the H$_2$ molecule in the complex of  He(1s2s,$^3$S$_1$) ($\equiv$ He$^{\ast}$) with H$_2$, in the dynamics in sub-kelvin regime. To elucidate physical phenomena presented in their novel experiment, the authors of Ref.~\cite{Klein:16} used the state-of-the-art first-principles calculations. The interaction energy of the complex comprised of the metastable helium atom and the hydrogen molecule was first calculated using the supermolecular approach employing the coupled cluster method with singles, doubles and perturbative triples (CCSD(T)), and is further denoted as $E_{\rm int}^{\rm CCSD(T)}$. The CCSD(T) method is known as the gold standard in quantum chemistry and in many cases provides values of properties accurate enough to predict experimental results~\cite{Vogels_2018}. However, in the case of He$^{\ast}$+H$_2$, the values of reaction rate coefficients calculated with the potential based on $E_{\rm int}^{\rm CCSD(T)}$ did not agree with experiment even qualitatively. Thus, the correction to the interaction energy, denoted here by $\delta E_{\rm int}^{\rm FCI}$, was calculated from the full configuration interaction (FCI) method and added to the $E_{\rm int}^{\rm CCSD(T)}$ energy. The agreement of the rate coefficients calculated from the $E_{\rm int}^{\rm CCSD(T)}$+$\delta E_{\rm int}^{\rm FCI}$ energies with the experimental values was improved but still qualitatively incorrect, since one additional resonance, not present in the experiment, was predicted for very low collision energies. To obtain quantitative agreement of the calculated rate coefficients with the experiment, the authors scaled the correlation part of the interaction energy by a factor of 1.004, that can be viewed as adding the correction $0.4\%E_{\rm int}^{\rm corr}$ to the $E_{\rm int}^{\rm CCSD(T)}$+$\delta E_{\rm int}^{\rm FCI}$ energy, suggesting that the basis set incompleteness was responsible for the inaccuracy of the rigorous {\em ab~initio} interaction energy surface.

The motivation of the present studies was to find a reason why the {\em ab~initio} interaction energy surface, even obtained at the FCI level of theory, was not accurate enough to precisely reproduce the experimental results. Here, we show that the missing piece of the puzzle is the nonrigidity of the H$_2$ molecule, neglected in the theoretical study of Ref.~\cite{Klein:16}. It has been recently shown that taking into account the monomer nonrigidity effects is necessary to obtain very precise agreement with the spectroscopic or scattering experiments~\cite{Jeziorska:00, Jankowski:12, Jankowski:13, Faure:16, Stoecklin:17}. In the present paper, we demonstrate for the first time the importance of the monomer flexibility in low-energy molecular anisotropic collisions. Only by inclusion of vibrations of the molecule in description of the complex, we are able to correctly predict results of very subtle scattering experiments with no fine-tuning to the experimental data whatsoever. Moreover, we present how to incorporate, in a simple and effective way, the nonrigidity effects into theoretical studies for any colliding molecules.

\section*{Results and discussion}

Theoretical consideration of our problem can be divided into two steps: first, the preparation of the most reliable interaction energy surface and second, scattering calculations. Since the complex comprises the metastable atom He$^{\ast}$ and the diatomic molecule H$_2$, the positions of the nuclei can be described by three coordinates: the distance $R$ between He$^{\ast}$ and center of mass (COM) of H$_2$, the angle $\theta$ between the H$_2$ bond and the COM--He$^{\ast}$ direction, and the distance $r$ between the hydrogen nuclei. Thus, in principle one has to use the three variables $(R,\theta,r)$ to parametrize the interaction energy surface and next to perform the scattering calculations. However, even for relatively small atom--diatom systems the full-dimensional treatment is rather rare. In most applications, the quantum scattering calculations are performed within the rigid-rotor approximation, i.e., assuming that the molecules in the complex are rigid. This widely used approximation is well physically motivated, since the internal vibrations of interacting molecules are much more energetic than the intermolecular modes~\cite{Jeziorska:00}. Usually, in the rigid-rotor calculations the rigid monomer interaction energy surfaces are used, obtained from {\em ab initio} calculations for the monomers with fixed geometries. Such calculations have been employed also in Ref.~\cite{Klein:16} to study the PI reaction of He$^\ast$+H$_2$. However, it has been shown very recently~\cite{Faure:16} that if in the rigid-rotor scattering calculations one uses the interaction energy averaged over the vibrations of the monomers, the results are closer to the full-dimensional calculations and experimental data. It has been demonstrated that the vibrationally averaged surfaces perform better than the rigid-monomer ones also in predictions of other physical properties, like rovibrational spectra~\cite{Jeziorska:00, Jankowski:08a, Jankowski:08b, Jankowski:12, Jankowski:13} or virial coefficients~\cite{Garberoglio:14, Garberoglio:17}. A main drawback of such approximation is that, in principle, to obtain the vibrationally averaged surface, one has to know the corresponding full-dimensional one. To avoid construction of a full-dimensional surface and minimize the number of geometries for which {\em ab initio} calculations have to be performed, we use the method developed in Ref.~\cite{Jankowski:04}. The details of this method are presented in the Methods section. Thus, in the current study we try to capture the nonrigidity effect by using the interaction energy surface averaged over vibrations of H$_2$ and rigid-rotor scattering calculations. We have to emphasize that although the averaged surface depends only on the intermolecular coordinates $(R,\theta)$, it cannot be regarded as the rigid one, since to obtain it one has to perform {\em ab initio} calculations for various values of the intramolecular distance $r$. Such a surface includes in an effective way information about the nonrigidity effects of the complex.

The most common technique to obtain PI rate coefficients is to use the complex potential in which the imaginary part describes the losses due to the ionization process~\cite{Miller72, Siska93}. With such potential one can solve the Schr\"odinger equation for the nuclear coordinates, for instance, using the close-coupling scattering method. More recently, two of us developed a new approach based on adiabatic theory and scattering theory for cold collision experiments~\cite{Pawlak_JCP_2015, Pawlak_JPCA_2017} dubbed as adiabatic variational theory (AVT). AVT together with the dedicated new closed-form expression for PI rate coefficients~\cite{Pawlak_JCQC_2018} provides a relatively simple method to implement and was used in our investigation. This technique, where the diatom is treated as a rigid rotor, allows to uncouple the rotations of the diatom and the complex from the atom--molecule separation.

\begin{figure}
\begin{center}
\includegraphics[width=\columnwidth]{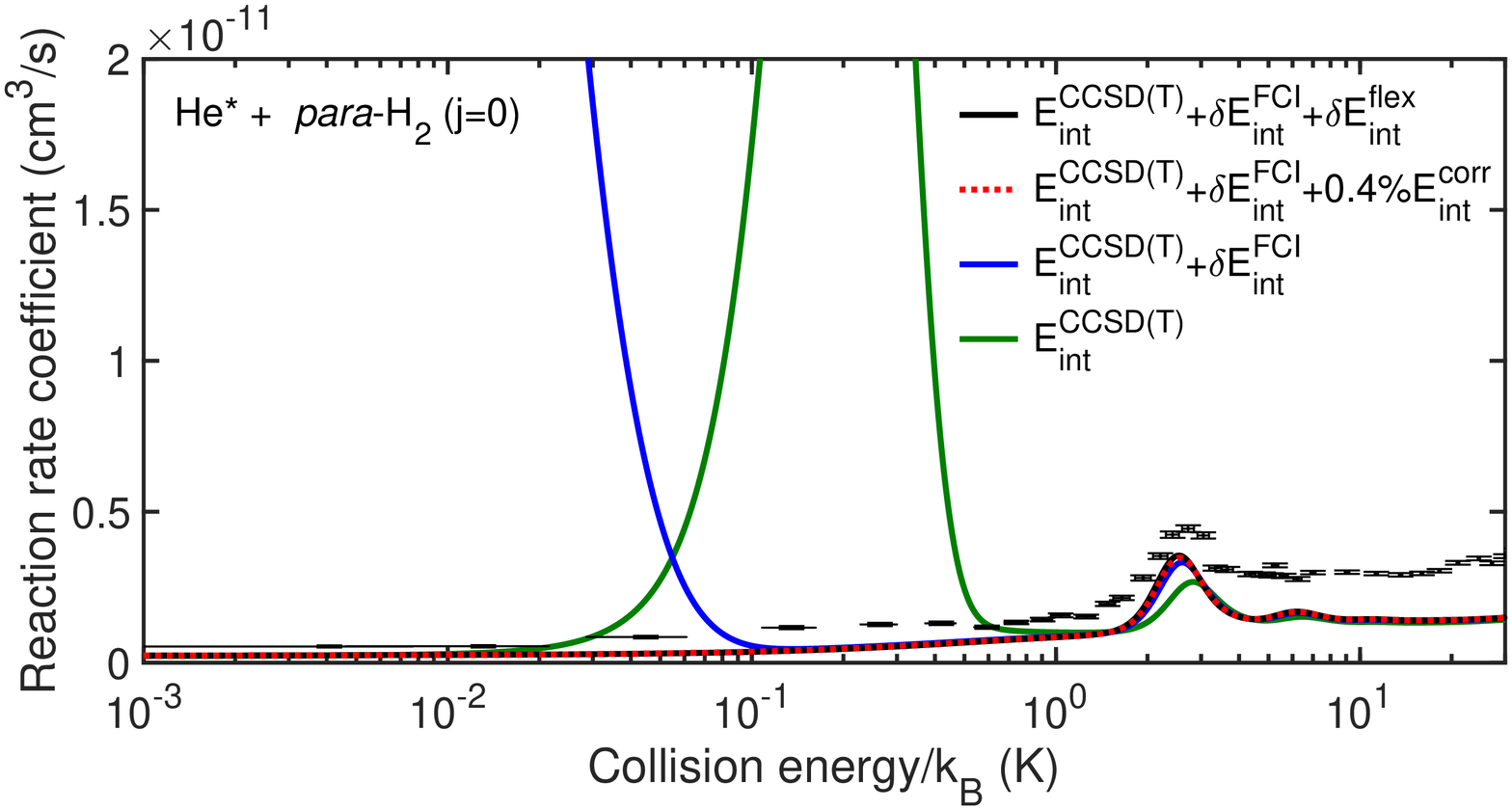}\\
\includegraphics[width=\columnwidth]{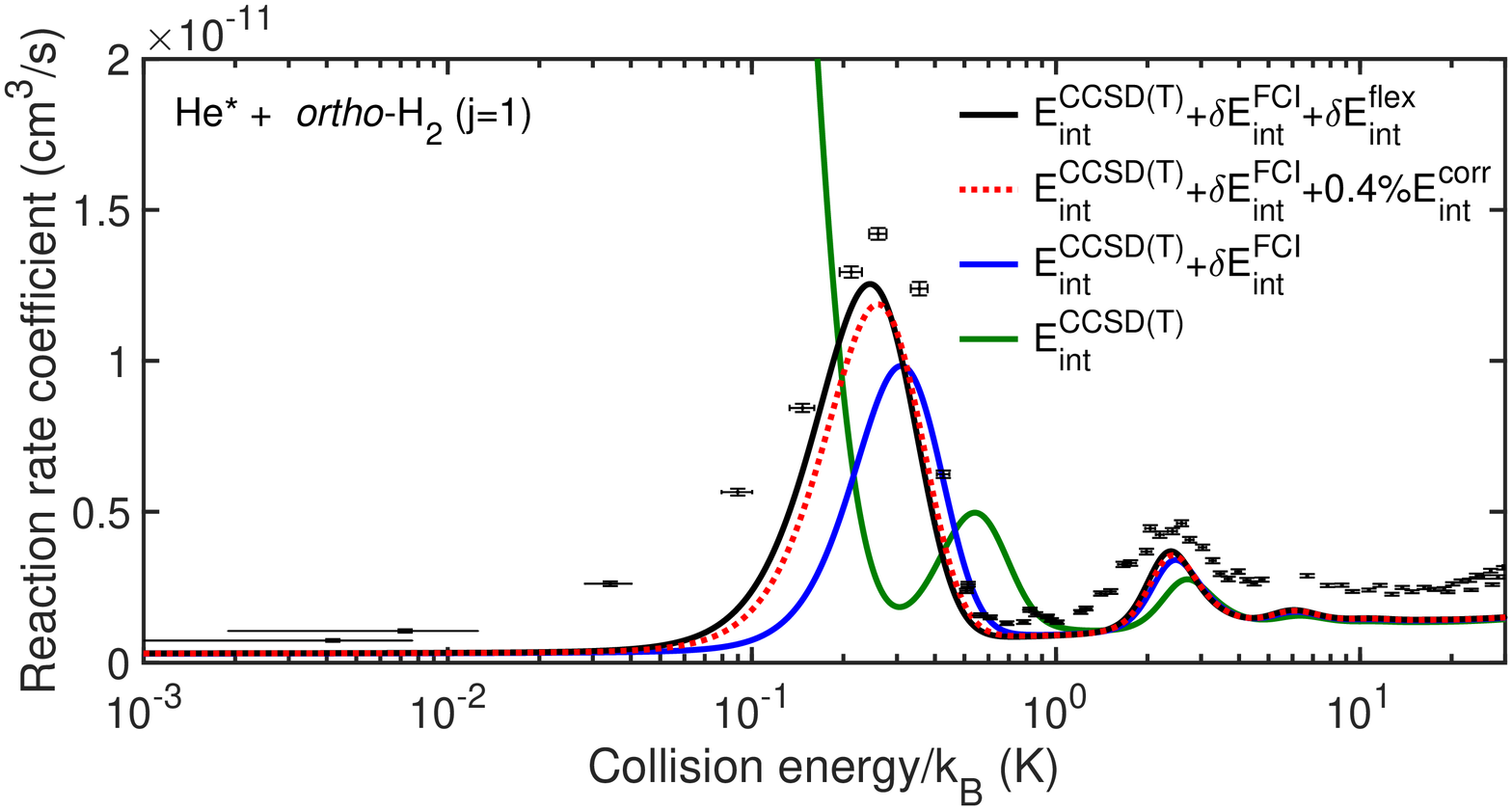}
\caption{Reaction rate coefficients of He(1s2s,$^3$S$_1$) with {\it para}-H$_2$ in the ground rotational state ($j=0$) [upper panel] and {\it ortho}-H$_2$ in the first excited rotational state ($j = 1$) [lower panel] with respect to relative energy (in K) between the colliding subsystems. The theoretical rate coefficients have been calculated based on the four interaction potentials:
(a) obtained at the CCSD(T) level of theory, $E^{\rm CCSD(T)}_{\rm int}$,
(b) the CCSD(T) one with the FCI correction, $E^{\rm CCSD(T)}_{\rm int}+\delta E^{\rm FCI}_{\rm int}$, 
(c) the CCSD(T) + $\delta$FCI surface with 0.4\% of the correlation energy added, $E^{\rm CCSD(T)}_{\rm int}+\delta E^{\rm FCI}_{\rm int}+ 0.4\% E^{\rm corr}_{\rm int}$, 
and (d) the CCSD(T) + $\delta$FCI surface with the correction describing the effect of the hydrogen molecule nonrigidity on the interaction energy, $E^{\rm CCSD(T)}_{\rm int}+\delta E^{\rm FCI}_{\rm int} + \delta  E^{\rm flex}_{\rm int}$.
The results have been convoluted with the experimental energy spread. Neither scaling nor fitting parameters have been used in the calculations. The experimental values are taken from Ref.~\cite{Klein:16} (black points with error bars). Theoretical results are slightly below the experimental data, however, the latter are burdened with systematic normalization error larger than the vertical discrepancy. 
} \label{fig_flex}
\end{center}
\end{figure}

\begin{figure}
\begin{center}
\includegraphics[width=\columnwidth]{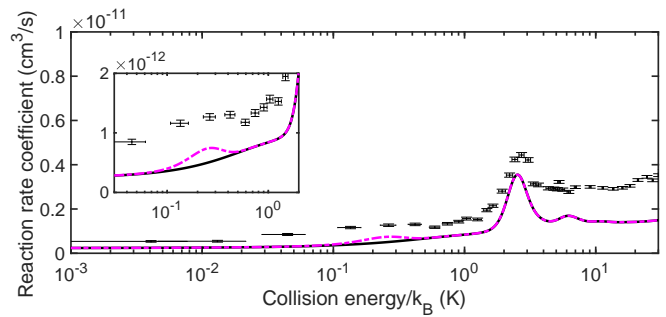}
\caption{Reaction rate coefficients of He(1s2s,$^3$S$_1$) with H$_2$ with respect to relative energy (in K) between the colliding subsystems. The theoretical rate coefficients have been calculated based on the interaction potential obtained using CCSD(T) with additional corrections resulting from FCI and inclusion of the effect of the hydrogen molecule nonrigidity. The black solid curve represents results for $100\%$ of {\it para}-H$_2$ ($j=0$), whereas the magenta dashed--dotted curve represents results for the mixture of $98\%$ of {\it para}-H$_2$ ($j=0$) and 2\% of {\it ortho}-H$_2$ ($j=1$). The inset exhibits the region with a small bump in millikelvin regime. The experimental data are shown as black points with error bars~\cite{Klein:16}. 
} \label{fig98}
\end{center}
\end{figure}

In Fig.~\ref{fig_flex} we present the reaction rate coefficients of He$^\ast$ with {\it para}-H$_2$ in the ground rotational state ($j=0$) and {\it ortho}-H$_2$ in the first excited rotational state ($j=1$). The strong effect of anisotropy on the resonant structure was discussed in details in Refs.~\cite{Klein:16, Pawlak_JPCA_2017}, however it is worth mentioning that in the interaction between He$^\ast$ and {\it para}-H$_2$ ($j=0$) only the isotropic part of the potential is probed, because the wavefunction of the molecular hydrogen in the lowest rotational state is spherically symmetric. When the interacting H$_2$ molecule is in the $j=1$ rotational state, the leading term of anisotropy of the potential contributes directly into the effective interaction and can firmly affect the positions of resonances. As demonstrated in Ref.~\cite{Pawlak_JPCA_2017}, by excluding from calculations the orientation-dependent part, the low-temperature resonance at collision energy of $k_B\times 0.27$~K is missed in the theoretical results. It shows that this peak, as opposed to the peak at 2.4~K, arises totally from the anisotropic interaction. It takes place when the excited helium atom collides with {\it ortho}-H$_2$ ($j=1$). Hence, the rate structures are  entirely different for different rotational states of the hydrogen molecule.

The evidence of monomer nonrigidity effects in cold anisotropic collisions is demonstrated in Fig.~\ref{fig_flex}. It is clearly seen that the results  with the interaction potential from the CCSD(T) method supplemented by the FCI correction are still not satisfactory. We found out that the discrepancy between the reported theoretical and experimental data is not due to the incompleteness of the used basis set as the authors of Ref.~\cite{Klein:16} stated, but due to the assumed stiffness of the molecule. By adding the correction corresponding to the flexibility of the diatom, termed as $\delta E^{\rm flex}_{\rm int}$, where the vibrations are averaged, we obtained an excellent agreement with the measurements over the whole range of temperatures. The nonrigidity correction shifts the energy of the resonance at 0.27~K to the position matching the experimental data. Also the magnitude of the rate coefficient curve is about 25\% larger than the one calculated without the nonrigidity correction. These results are very similar to those obtained in Ref.~\cite{Klein:16} with the artificial $0.4\%$ increase of the correlation part of the interaction energy. Note that in our entire calculations we did not apply any scaling or fitting parameters as well as we did not shift the final results to adjust to the experiment. Our results are slightly below the experimental ones, but the latter have been normalized to the absolute scale according to thermal rate observations at 300~K, see the Methods in Refs.~\cite{Ed_NatChem_2015, Klein:16}, and this procedure introduced a systematic error to the experimental data much larger than the vertical discrepancy. In the Supplementary Information we provide a figure corresponding to Fig.~\ref{fig_flex}, with the rate coefficient curves shifted by a constant value to match the normalized experimental data at the collision energy around $k_B\times 2.4$~K, as it was done in Ref.~\cite{Klein:16}. After such additional ``normalization'', the theoretical resonance structure around 0.27~K perfectly agrees with the experiment.

In the experimental data for {\it para}-H$_2$, given in the upper panel of Fig.~\ref{fig_flex}, one can see a very flat bump around the collision energy of $k_B\times 0.3$~K, not discussed in Ref.~\cite{Klein:16}. On the basis of our theoretical considerations we can explain an origin of that feature. One of the beams used in the Narevicius' experiment is formed of the {\it para}-H$_2$ molecules. However, a purity of {\it para}-H$_2$ was limited to $98\%$~\cite{Klein:16}. Therefore, we have added $2\%$ of the reaction rate coefficients of {\it ortho}-H$_2$ ($j=1$) given in the lower panel of Fig.~\ref{fig_flex} to $98\%$ of the reaction rate coefficients of {\it para}-H$_2$ ($j=0$) given in the upper panel of Fig.~\ref{fig_flex}; the resulting rates are presented in Fig.~\ref{fig98}. Now one can see, in the inset of this figure, that the bump appears on the theoretical curve, properly predicting the position and the shape of its experimental counterpart.

\section*{Conclusions}

Low-energy collision experiments allow to directly and clearly reveal the true nature of interaction. Only at a high level of theory we are able to predict or confirm experimental data as well as understand quantum phenomena in chemical reactions. In this paper, we have demonstrated a significant role of monomer nonrigidity effects on the position and intensity of scattering resonances in anisotropic cold molecular collisions. We have investigated excited metastable helium atoms colliding with hydrogen molecules in the temperature range from a few dozen kelvins to 1~millikelvin. We have provided the most accurate interaction energy surface that can be treated as the reference one for novel quantum chemistry methods for molecular systems in the resonance state. The calculated rate coefficients are in remarkable agreement with measurements~\cite{Klein:16}. We have demonstrated that the discrepancy between the experimental and theoretical results discussed in Ref.~\cite{Klein:16} is due to the assumption that the H$_2$ molecule is rigid. We have complemented the theoretical description of the interaction energy of the complex by inclusion of its dependence on the flexibility of the molecule. It shows that the approach beyond the commonly used rigid-rotor approximation is indispensable even when rigorous state-of-the-art computations are performed at the FCI level of theory. The correction to the interaction energy due to the nonrigidity of the monomer is a few times larger than the uncertainties caused by basis-set incompleteness or generated by the Born--Oppenheimer approximation. Our finding is not limited to a specific complex considered here nor restricted to the PI reaction process. We believe that it can be vitally important in precisely controlled cold chemistry, where quantum effects in chemical reactions dominate.

\section*{Methods}

\subsection*{Vibrationally averaged surface}

To construct an interaction energy surface averaged over the vibrations of the monomers, $\langle V \rangle$, one can take advantage of the fact that the molecules in the complex preserve their identity and frequencies of internal vibrations are much higher than those of the intermolecular modes, and perform averaging over the intramonomer coordinates similar to the adiabatic approximation in the electronic-structure theory~\cite{Jeziorska:00}. Since we are interested in the He$^{\ast}$+H$_2$ complex, let us limit our consideration to the atom-diatom case. If the Jacobi coordinates are used, the interaction energy surface $V(R,\theta,r)$ can be represented as the truncated Taylor expansion around some reference geometry $r_c$~\cite{Jankowski:04}
\begin{equation}
V_{\rm TE}(R,\theta,r)= f_0(R,\theta)+f_1(R,\theta)(r-r_c)
+\frac{1}{2}f_2(R,\theta)(r-r_c)^2,
\label{Vexpand}
\end{equation}
where $f_0 (R,\theta) = V(R,\theta,r_c)$, $f_1 (R,\theta) = \frac{\partial V(R,\theta,r)}{\partial r}\big|_{r=r_c}$, and $f_2 (R,\theta)= \frac{\partial^2 V(R,\theta,r)}{\partial r^2}\big|_{r=r_c}$. The higher order terms can be neglected if only modest deformations of the monomer are allowed, as those corresponding to a few lowest vibrational states of the H$_2$ molecule. The $f_0$ function is in fact the rigid monomer two-dimensional surface calculated for the H--H separation equal $r_c$, while the remaining terms account for the nonrigidity effects, i.e., the dependence of the surface on the intramolecular coordinate. The potential $V_{\rm TE}$ from Eq.~(\ref{Vexpand}) can be easily averaged over the $v$ vibrational state of the monomer and the resulting formula reads
\begin{eqnarray}
\langle V_{\rm TE} \rangle (R,\theta) 
&=& f_0(R,\theta)+f_1(R,\theta)\left(\langle r \rangle_v - r_c\right) \nonumber \\
&& +\frac{1}{2}f_2(R,\theta)
  \left(\langle r^2 \rangle_v - 2 \langle r\rangle_v r_c + r_c^2 \right).
\label{VTEavB}
\end{eqnarray}
We can use the $\langle V_{\rm TE} \rangle$ surface as an approximation to the $\langle V \rangle$ one, i.e., assume $\langle V \rangle \approx \langle V_{\rm TE} \rangle$. The values of $\langle r\rangle_v$ and $\langle r^2\rangle_v$ can be calculated from the theoretical properties of the monomer or even estimated from the empirical spectroscopic constants. The surfaces averaged according to the approximation of Eq.~(\ref{VTEavB}) and the corresponding formula for the diatom-diatom case, were used in the rigid-rotor dynamical calculations, both the bound-states and the scattering, and provided the results in excellent agreement with the experimental ones~\cite{Jankowski:08a, Jankowski:08b, Jankowski:12, Jankowski:13, Faure:16}. In practical applications, there is no need to know the surface $\langle V_{\rm TE} \rangle$ for any values of the $(R,\theta)$ coordinates, but it would be enough to calculate it on the grid points, for instance the ones used in the scattering calculations. For each grid point in $(R,\theta)$, we can compute the interaction energy $f_0$ and the numerical values of the $f_1$ and $f_2$ derivatives. Of course, one can also calculate $\langle V_{\rm TE} \rangle$ for a given grid of geometries and then fit an analytical function to obtain the surface. From the expression (\ref{VTEavB}), one can see a very useful feature, namely that the values of $f_1$ and $f_2$  may be calculated on a different level of theory than $f_0$, e.g., the leading term $f_0$ on the highest possible level, and the values of $f_i$ defining the higher order terms, specifying the dependence of $V_{\rm TE}$ on $r$, may be calculated at a somehow lower level of theory. Since the first and second order terms are much smaller than the leading term $f_0$, such an additional approximation only slightly increases the uncertainty of $V_{\rm TE}$ or $\langle V_{\rm TE} \rangle$, whereas it may significantly reduce the computational effort required. Such a strategy has been applied, for instance, to the H$_2$--CO complex in Refs.~\cite{Jankowski:12, Jankowski:13} and led to the very accurate rovibrational spectra.

To prepare the vibrationally averaged surface for the He$^{\ast}$+H$_2$ complex, we have used the formula of Eq.~(\ref{VTEavB}) in the following way. The leading term $f_0$ was set to be equal to the rigid-monomer interaction energy of Ref.~\cite{Klein:16}. That interaction energy can be written as $E_{\rm int}^{\rm CCSD(T)}$+$\delta E_{\rm int}^{\rm FCI}$, using our notation, and was obtained as a sum of the interaction energy calculated at the CCSD(T) level and the FCI correction for the intramolecular distance 1.4487~bohr. Thus, we have to set $r_c$ to be equal to the same value to make our choice of $f_0$ consistent with Eq.~(\ref{VTEavB}). Since we consider the ground vibrational state of H$_2$, $v=0$, the values of  $\langle r\rangle_0$ and $\langle r^2\rangle_0$ for $j=0$ were set to the round-off values from Ref.~\cite{Bubin:03} equal to 1.4487~bohr and 2.1270~bohr, respectively. The equality of $r_c$ and $\langle r\rangle_0$ is of course not accidental, because the interaction energy of Ref.~\cite{Klein:16} was calculated for the rigid H$_2$ molecule with the vibrationally averaged geometry. With such a choice of $r_c$, the first-order term in Eq.~(\ref{VTEavB}) vanishes and the nonrigidity effect is fully described by the second-order term. The values of $\langle r\rangle_0$ and $\langle r^2\rangle_0$ for the molecule in the first excited rotational state ($j=1$) are slightly different: 1.4509~bohr and 2.1334~bohr, respectively. Therefore, for this case the first-order term in Eq.~(\ref{VTEavB}) contributes to the nonrigidity correction.

In the standard, supermolecular approach the interaction potential is very difficult to obtain due to the fact that our potential is not the ground-state one and is coupled to the scattering state of the He+H$_2^+$+$e^-$ system. In Ref.~\cite{Hapka13} it was shown that using a carefully tailored start guess it is possible to converge the CCSD(T) interaction potential and that also the symmetry-adapted perturbation theory (SAPT)~\cite{Jeziorski_ChemRev_94, Hapka_2012, Zuchowski_2003} provides a good representation of the short-range potential. Nonetheless, it is very difficult to stabilize first and second derivatives of the interaction potential with respect to the nuclear coordinate motion in the supermolecular method as it inherently relies on subtraction of big numbers and loss of accuracy is unavoidable. We previously stated that the derivative of the interaction energy can be obtained at a lower level of theory and it still catches the essential physics. As a matter of fact, the application of SAPT greatly facilitates the calculation of nonrigidity effects. Since in SAPT the interaction energy is obtained directly from the wavefunction of monomers, the interaction energy is stable and very inexpensive. This is due to the fact that in SAPT we obtain the interaction energy {\em directly} as a sum of the perturbation theory terms in which the expansion parameter is the interaction potential between monomers. Here we use the interaction energy which is the sum of first two terms of perturbation series in the interaction operator between H$_2$ and helium analogously to Ref.~\cite{Hapka13}. The values of $f_1$ and $f_2$ were calculated from the four- and five-point central-difference formula, respectively, with $h=0.05$~bohr. For each grid point in $(R,\theta)$, the interaction energy was calculated for five separations $r_c+kh$, $k=-2,-1,0,1,2$. The SAPT calculations were carried out with the d-aug-cc-pVTZ basis set. We estimated that uncertainty of calculation of flexibility correction in the interaction well is underestimated by about 2\% for linear geometry and about 10\% for T-shape geometry. In absolute numbers these uncertainties are below 0.01~cm$^{-1}$. For discussion of uncertainty of $\delta E^{\rm flex}_{\rm int}$ see the Supplementary Information.

In Ref.~\cite{Klein:16}, the effect on the interaction energy surface beyond the Born--Oppenheimer approximation was neglected on the basis of the analysis of the value of the diagonal Born--Oppenheimer (DBO) correction~\cite{Handy:86} calculated at the minimum of the surface. Here we performed an extended analysis for two angular orientations, $\theta=0^{\circ}$ and $90^{\circ}$, several values of $R$, and the H$_2$ geometry fixed at $r_c$. The DBO correction to the interaction energy, $\delta E_{\rm int}^{\rm DBO}$, was obtained by subtracting from the DBO correction calculated for the complex the value calculated for the monomer at large separations of interacting species. The calculations were performed at the CCSD level of theory, with the aug-cc-pVTZ basis set augmented by the midbond functions. This $\delta E_{\rm int}^{\rm DBO}$ correction causes a small positive shift of the interaction energy, smaller than about 0.06~cm$^{-1}$ in the minimum region, i.e., five times smaller than the value of $\delta E^{\rm flex}_{\rm int}$ at this geometry. The values of $\delta E_{\rm int}^{\rm DBO}$ for some other geometries are given in the Supplementary Information. We have also added $\delta E^{\rm DBO}_{\rm int}$ to the $E^{\rm CCSD(T)}_{\rm int}+\delta E^{\rm FCI}_{\rm int} + \delta  E^{\rm flex}_{\rm int}$ energy and have performed the scattering calculations. The resulting positions of resonances are almost unaffected by the addition of the $\delta E^{\rm DBO}_{\rm int}$ correction, as can be seen in the figure presented in the Supplementary Information.

Finally, let us discuss the effect of basis set incompleteness. It is very difficult to estimate since the basis set extrapolation techniques are questionable to use in the present case, where the system is not in its ground state and the system concerned is essentially a resonance. The basis set used in the present calculations at the CCSD(T) level, aug-cc-pV6Z with bond functions, is already well saturated for the dispersion energy which dominates the interaction energy for the considered system. A computationally expensive increase of the basis set to aug-cc-pV7Z shifts the interaction energy in the global (linear) minimum by about $-0.022$~cm$^{-1}$, while for the local minimum at a T-shape geometry (for 10.85~bohr) the shift is about $-0.027$~cm$^{-1}$. Close to the inner turning points at 9.5~bohr, these values are $-0.034$ and $-0.045$~cm$^{-1}$, respectively. Unfortunately, one cannot perform an extrapolation to the complete basis set, since the resonant nature of the interaction affects the stability of {\it ab initio} calculations at the level of accuracy of the order of 0.01~cm$^{-1}$. Nonetheless, one should bear in mind that such an interaction energy should not be extrapolated in the usual sense but rather than that, stabilized. For this particular system, the coupling between the continuum and the bound state is weak, hence, the potential obtained by the standard quantum chemistry methods gives right answer for right reasons. An improvement of theory to go below the 0.01~cm$^{-1}$ accuracy is a formidable task and a new approach would be needed to address such demands. Since the real potential is dominated by the dispersion energy which is a variational quantity, the more complete basis set implies the deeper potential. However, if one proceeds from the aug-cc-pV6Z basis set to the aug-cc-pV7Z one, the shift of the interaction energy surface by 0.02--0.03~cm$^{-1}$ at the minimum separation is an order of magnitude smaller than in the case of the $\delta E^{\rm flex}_{\rm int}$ correction. Thus, the effect of the basis set incompleteness on the positions and intensities of the resonances is negligible in comparison to the effects caused by the $\delta E^{\rm flex}_{\rm int}$ correction and is also much smaller than the effect of the $\delta E_{\rm int}^{\rm DBO}$ correction.

\begin{figure}[t]
\begin{flushright}
\includegraphics[width=8.4cm]{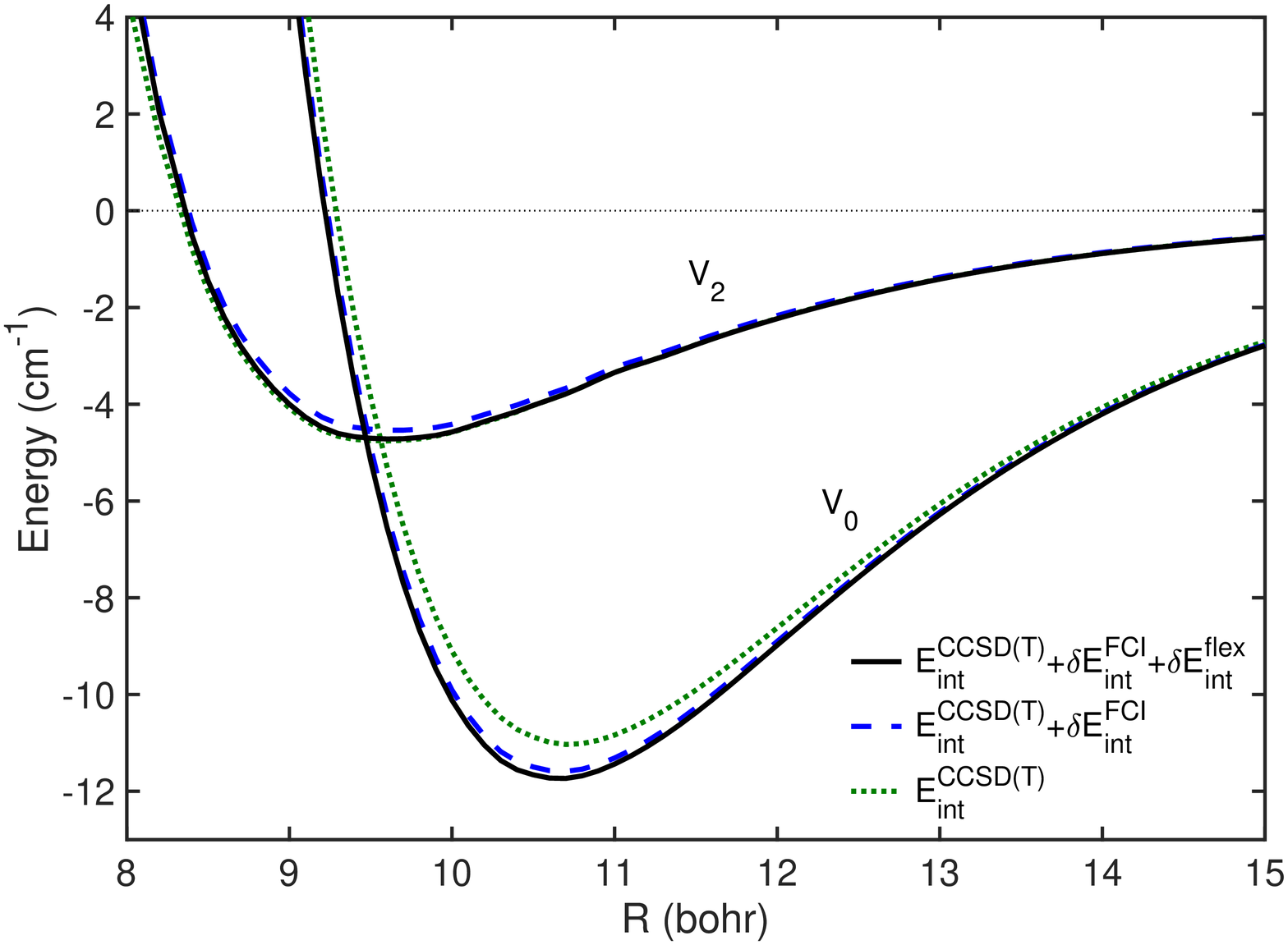}\\
\vspace{0.3cm}
\includegraphics[width=8.4cm]{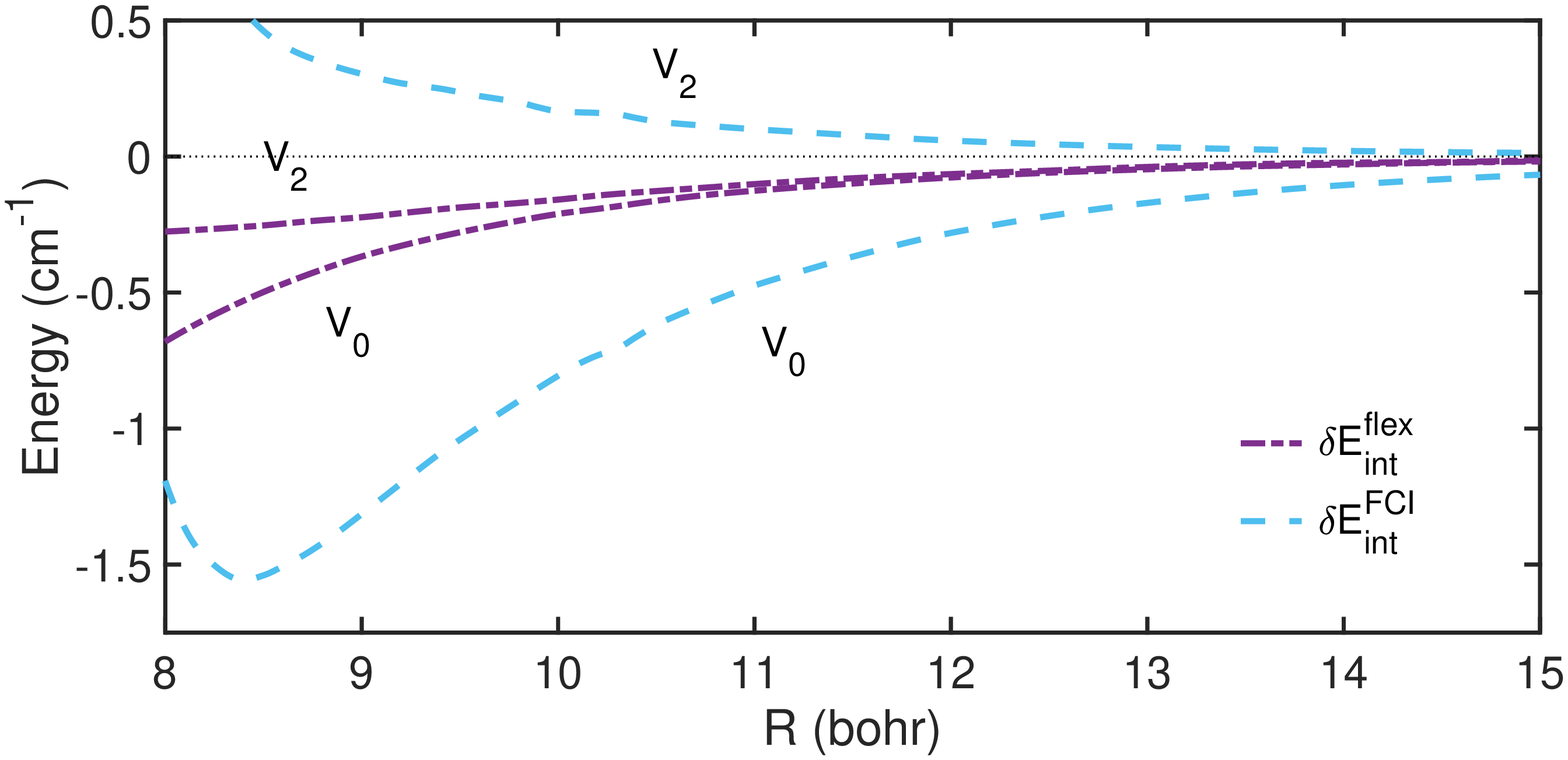}
\caption{Real part of the interaction energy components of the He$^{\ast}$+H$_2$ potential in various approximations. Upper panel: Isotropic, $V_0(R)$, and anisotropic, $V_2(R)$, radial interaction potential terms, obtained at a few levels of theory: CCSD(T) (in green), CCSD(T) plus the correction resulting from FCI (in blue), and CCSD(T) plus the FCI correction plus the correction due to the effect of the hydrogen molecule nonrigidity (in black). Lower panel: Values of the $\delta E^{\rm FCI}_{\rm int}$ and $\delta E^{\rm flex}_{\rm int}$ corrections corresponding to the $V_0$ and $V_2$ terms.
} \label{fig_pot}
\end{flushright}
\end{figure}

According to the works of Refs.~\cite{Klemperer77,Dubernet_Hutson_JCP_1994}, we may expand the vibrationally averaged surface $\langle V\rangle$ in Legendre polynomials
\begin{equation}\label{pot_eq}
\langle V\rangle (R,\theta) = \sum_\eta V_\eta (R)P_\eta(\cos\theta).
\end{equation}
For the collision of an atom with a homonuclear diatomic molecule the index $\eta$ is even due to symmetry reasons, $\langle V\rangle (R,-\theta) = \langle V\rangle (R,\theta)$. In other words, terms for odd $\eta$ vanish. Thus, the two leading terms are given by $V_0(R)$ and $V_2(R)(3\cos^2\theta-1)/2$, where $V_0(R)$ and $V_2(R)$ are radial isotropic and anisotropic interaction potential terms, respectively.

It should be emphasized that the considered complex is not in the bound state but in the resonance one. The total potential energy surface is above the ionization threshold, thus the electronic state of the He(1s2s,$^3$S$_1$)+H$_2$ system is embedded in the continuum of the He(1s$^2$,$^1$S$_1$)+H$_2^{+}$+$e^{-}$ system. Consequently, the total potential energy surface has to be complex where the imaginary part represents the ionization width (inverse lifetime). Two new approaches have been lately developed for PI widths: one is based on Fano-algebraic diagramatic construction method~\cite{Vitali_JCP_2018}, the next one uses the stabilization method with an analytical continuation via the Pad\'{e} approximant~\cite{Landau_JCPA_2016, Debarati_JCQC_2017}. In our studies, we took the imaginary part of $V_0$ and $V_2$ from Ref.~\cite{Debarati_JCQC_2017}, obtained by the latter technique.

The calculated isotropic, $V_0$, and anisotropic, $V_2$, radial interaction potential terms obtained on three levels of theory are presented in the upper panel of Fig.~\ref{fig_pot}. One can see that adding the $\delta E_{\rm int}^{\rm FCI}$ correction to $E_{\rm int}^{\rm CCSD(T)}$ apparently changes both $V_0$ and $V_2$, but in opposite directions: $V_0$ becomes deeper and $V_2$ slightly shallower. The subsequent addition of $\delta E_{\rm int}^{\rm flex}$ to the $E_{\rm int}^{\rm CCSD(T)}$+$\delta E_{\rm int}^{\rm FCI}$ interaction energy makes the resulting $V_0$ even deeper, whereas for the $V_2$ term the effect of $\delta E_{\rm int}^{\rm flex}$ almost cancels the effect of $\delta E_{\rm int}^{\rm FCI}$. The lower panel of Fig.~\ref{fig_pot} presents how the $\delta E_{\rm int}^{\rm FCI}$ and $\delta E_{\rm int}^{\rm flex}$ corrections to the interaction energy enter the $V_0$ and $V_2$ terms of the interaction potential.

\subsection*{Adiabatic variational theory}

To solve the Schr\"odinger equation for the complex with the previously prepared interaction potential and consequently to calculate rate coefficients, we used the AVT approach that has been developed for cold atom--molecule collision experiments~\cite{Pawlak_JCP_2015, Pawlak_JPCA_2017}. This technique has been recently successfully applied for the He(1s2p,$^3$P$_2$)+H$_2$ system~\cite{Bhattacharya_2019}. It enables one to reduce the complexity of the problem enhancing the computational performance without losing physical essence. Within AVT, we represent the potential (\ref{pot_eq}) in a basis set consisting of many angular functions. The single angular function in our case is a product of two spherical harmonics --- one is responsible for the description of rotations of the molecule, whereas the other one for the description of rotations of the whole complex. Such a matrix constructed for a given intermolecular distance $R$ needs to be diagonalized providing a set of eigenvalues. The process has to be repeated for different values of $R$. The obtained eigenvalues, after ordering, create so-called adiabats (effective potentials) depending on $R$. Then, from the practical reasons all adiabats are shifted to get asymptotes at zero. Therefore, for each of them the dissociation threshold is at zero. Next, we solve the one-dimensional Schr\"odinger equation many times, each time with a different adiabat treating $R$ as a variable. Here, any technique can be used, e.g., by spanning the wave function space in a basis of trial functions~\cite{Pawlak_PRA_2011, Pawlak_PRA_2014-Rydberg}. Finally, we apply the simple and easy to implement formula for reaction rate coefficients that has been derived based on AVT and non-Hermitian scattering theory~\cite{Pawlak_JCQC_2018}. Only the information about complex eigenenergies, the reduced mass of the atom--diatom system, and the rotational state of the molecule are required. In calculations, we used 21 partial waves corresponding to end-over-end angular momenta of the complex ($l=0,1,...,20)$. The Schr\"odinger equation was solved by the DVR with the box size of 500~bohr and with 2000 basis sine-functions. The results are fully converged with respect to the number of partial waves and of basis functions. The calculated reaction rate coefficients are convoluted over the experimental collision energy spread (8~mK). Neither scaling nor fitting parameters have been used in our calculations.

\section*{Acknowledgements}
We would like to thank Debarati Bhattacharya for providing the imaginary part for radial interaction potentials.

This research was financially supported by the National Science Centre, Poland 
(Grant No. 2016/23/D/ST4/00341 [MP], Grant No. 2015/19/B/ST4/02707 [PS\.{Z}], and Grant No. 2017/25/B/ST4/01300 [PJ]).

\section*{Competing financial interests}
The authors declare no competing financial interests.

\end{document}


\title{Supplementary Information for:\\ Nonrigidity effects --- a missing puzzle piece in the description of low-energy anisotropic molecular collisions}

\author{Mariusz Pawlak}
\email{e-mail: teomar@chem.umk.pl}
\affiliation{Faculty of Chemistry, Nicolaus Copernicus University in Toru\'n, Gagarina~7, 87-100~Toru\'n, Poland}
\author{Piotr S. {\.Z}uchowski}
\email{e-mail: pzuch@fizyka.umk.pl}
\affiliation{Faculty of Physics, Astronomy and Informatics, Nicolaus Copernicus University in Toru\'n, Grudzi\k{a}dzka~5, 87-100~Toru\'n, Poland}
\author{Nimrod Moiseyev}
\affiliation{Schulich Faculty of Chemistry and Department of Physics, Technion--Israel Institute of Technology, Haifa 32000, Israel}
\author{Piotr Jankowski}
\affiliation{Faculty of Chemistry, Nicolaus Copernicus University in Toru\'n, Gagarina~7, 87-100~Toru\'n, Poland}

\renewcommand{\thepage}{S\arabic{page}}
\renewcommand{\thetable}{S\arabic{table}}
\renewcommand{\thefigure}{S\arabic{figure}}

\maketitle

In Table~\ref{tab1} we present, for selected geometries of the complex, the values of the leading part of the interaction energy $E_{\rm int}^{\rm CCSD(T)}$ obtained at the CCSD(T) level of theory and the values of various corrections to this energy. One can see that for the geometry close to the global minimum of the interaction energy surface, for $\theta=0^\circ$ and $R=10.5$ bohr, the value of $\delta E_{\rm int}^{\rm flex}$ is equal to $-0.29$~cm$^{-1}$ and, although it seems to be small in the absolute scale, amounts to about 2\% of the $E_{\rm int}^{\rm CCSD(T)}$+$\delta E_{\rm int}^{\rm FCI}$ interaction energy. For the same distance and $\theta=90^\circ$, $\delta E_{\rm int}^{\rm flex}$ amounts to 1\% of the total energy. Thus, the $\delta E_{\rm int}^{\rm flex}$ correction slightly change the anisotropy of the potential. It is also interesting that the ratio of $\delta E_{\rm int}^{\rm flex}$ to $\delta E_{\rm int}^{\rm FCI}$ is significantly different for $\theta=0^\circ$ and $90^\circ$, and amounts to 0.60 and 0.15, respectively, that shows that the anisotropy of these two corrections is completely different. The ratio is almost constant for the whole range of values of $R$ at the same values of $\theta$. The most important comparison one can draw from Table~\ref{tab1} is between the $\delta E_{\rm int}^{\rm flex}$ and $0.4\%E_{\rm int}^{\rm corr}$ corrections. For $\theta=90^\circ$ they are almost equal, while for $\theta=0^\circ$ the values of $\delta E_{\rm int}^{\rm flex}$ are about two times larger than the $0.4\%E_{\rm int}^{\rm corr}$ ones. Nevertheless, there is no doubts that the $0.4\%E_{\rm int}^{\rm corr}$ correction introduced in Ref.~\cite{Klein:16} to reproduce the experimental rate coefficients, can be recognized, on the basis of our investigation, as the result of the nonrigidity effect of H$_2$ on the interaction energy.

\begin{figure}[t!]
\begin{center}
\includegraphics[width=11cm]{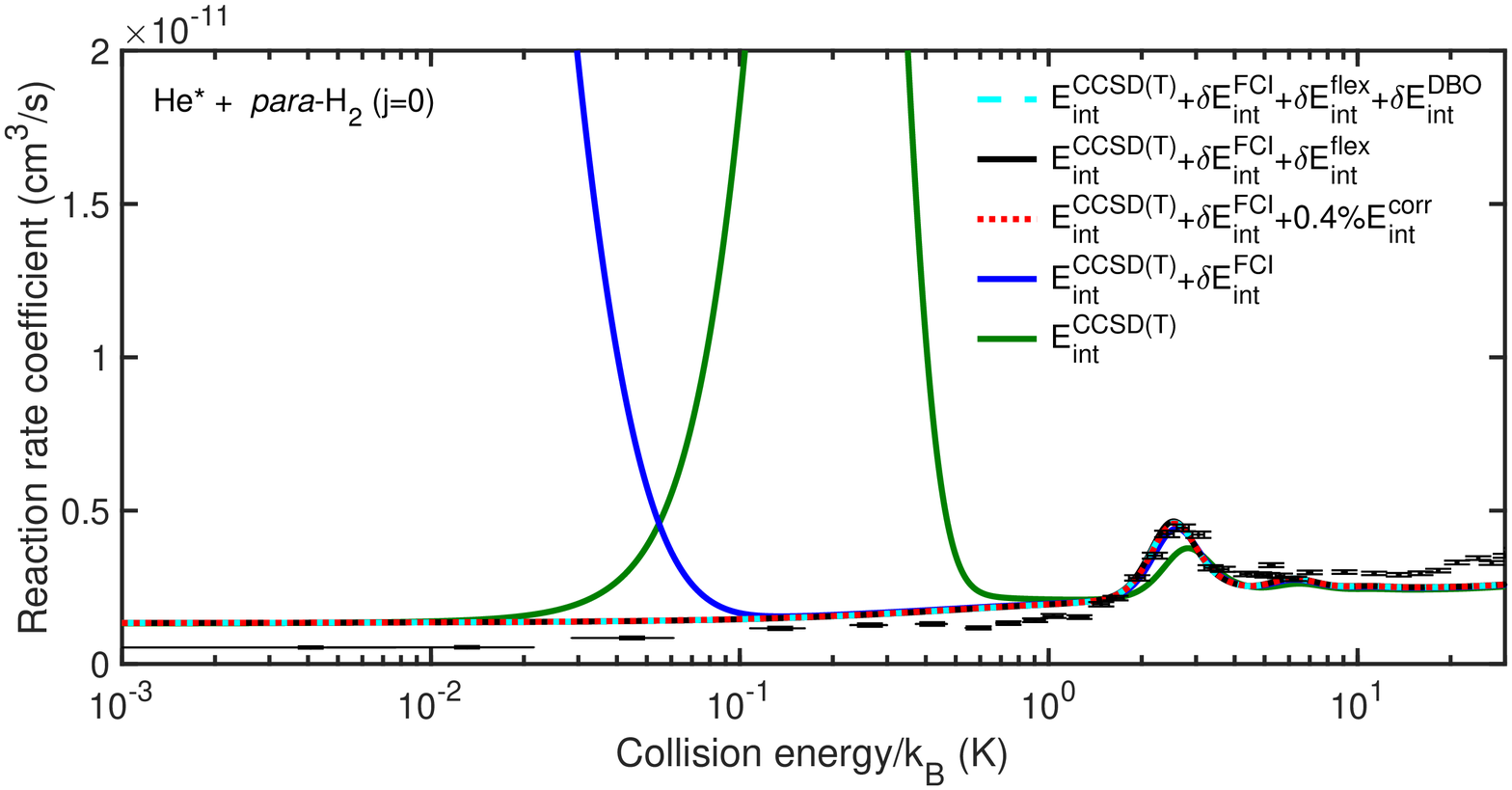}\\
\includegraphics[width=11cm]{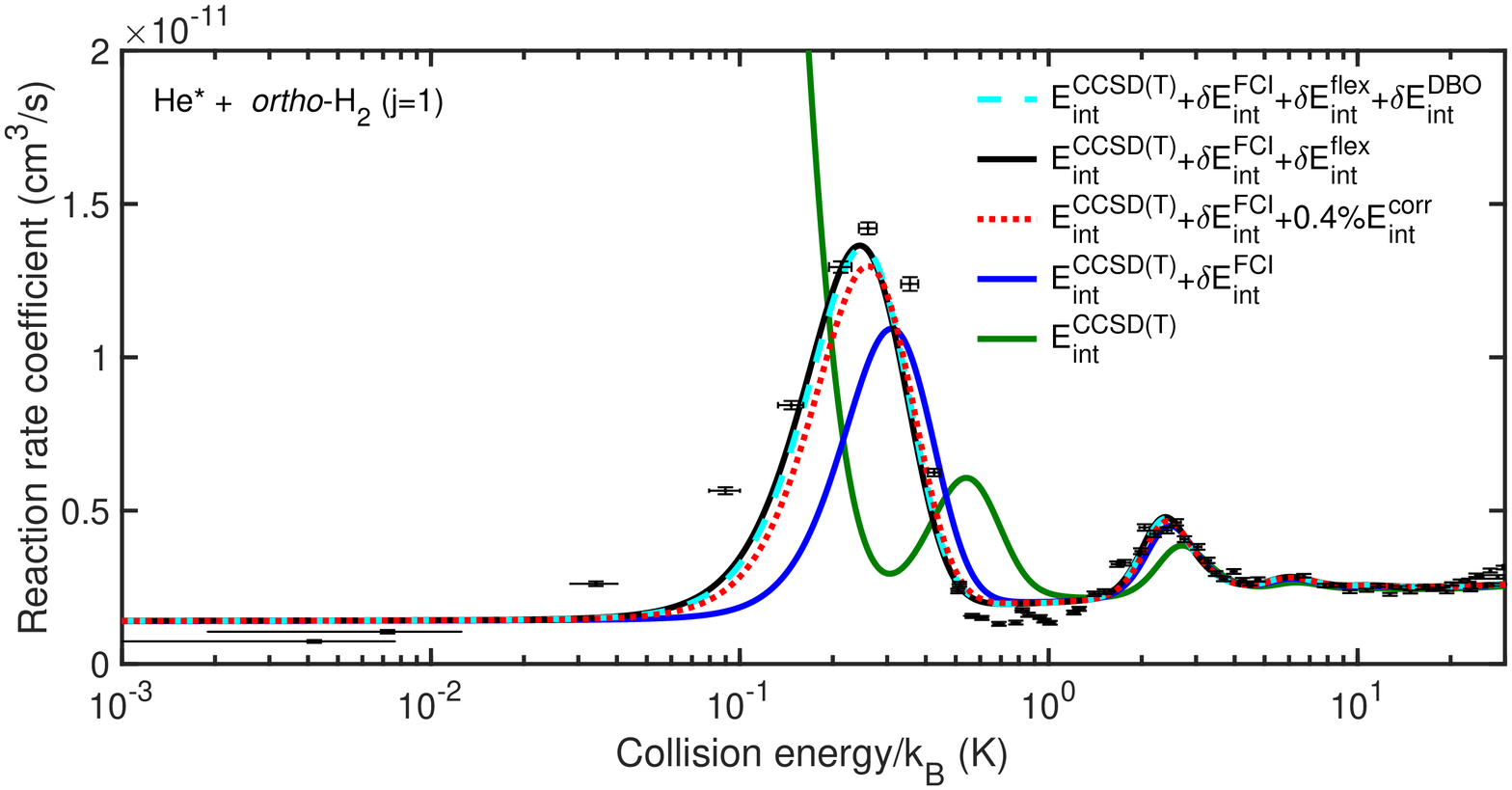}
\caption{Reaction rate coefficients of He(1s2s,$^3$S$_1$) with {\it para}-H$_2$ in the ground rotational state ($j=0$) [upper panel] and {\it ortho}-H$_2$ in the first excited rotational state ($j = 1$) [lower panel] with respect to relative energy (in K) between the colliding subsystems. The theoretical rate coefficients have been calculated based on the five interaction potentials: 
(a) obtained at the CCSD(T) level of theory, $E^{\rm CCSD(T)}_{\rm int}$,
(b) the CCSD(T) one with the FCI correction, $E^{\rm CCSD(T)}_{\rm int}+\delta E^{\rm FCI}_{\rm int}$, 
(c) the CCSD(T) + $\delta$FCI surface with 0.4\% of the correlation energy added, $E^{\rm CCSD(T)}_{\rm int}+\delta E^{\rm FCI}_{\rm int}+ 0.4\% E^{\rm corr}_{\rm int}$, 
(d) the CCSD(T) + $\delta$FCI surface with the correction describing the effect of the hydrogen molecule nonrigidity on the interaction energy, $E^{\rm CCSD(T)}_{\rm int}+\delta E^{\rm FCI}_{\rm int} + \delta  E^{\rm flex}_{\rm int}$, 
and (e) the CCSD(T) + $\delta$FCI + $\delta$flex surface with the diagonal Born--Oppenheimer correction, $E^{\rm CCSD(T)}_{\rm int}+\delta E^{\rm FCI}_{\rm int} + \delta  E^{\rm flex}_{\rm int} + \delta  E^{\rm DBO}_{\rm int}$.
The results have been convoluted with the experimental energy spread. Neither scaling nor fitting parameters have been used in the calculations. The experimental values are taken from Ref. \cite{Klein:16} (black points with error bars). Theoretical results are shifted up by a constant value to match the normalized experimental data at the collision energy around 2.4~K.
} \label{fig_flex}
\end{center}
\end{figure}

In Table~\ref{tab1} we present also the values of the diagonal Born--Oppenheimer (DBO) correction, $\delta E^{\rm DBO}_{\rm int}$, calculated at the CCSD level of theory, in the way described in the Method section of the paper. For the values of $R$ smaller than 10.0~bohr for $\theta=0^\circ$ and 9.0~bohr for $\theta=90^\circ$, we have problems to converge the calculations of $\delta E^{\rm DBO}_{\rm int}$ at the CCSD level, thus for small values of $R$ in the scattering calculations, we have used the values of  $\delta E^{\rm DBO}_{\rm int}$ extrapolated from the region of $R$ greater than or equal to 10.0~bohr and 9.0~bohr for $\theta=0^\circ$ and $90^\circ$, respectively. From Table~\ref{tab1} one can see that for geometries close to the geometry of the minimum, $\theta=0^\circ$ and $R=10.5$~bohr, the value of $\delta E^{\rm DBO}_{\rm int}$ amounts to 20\% of $\delta E^{\rm flex}_{\rm int}$. The $\delta E^{\rm DBO}_{\rm int}/\delta E^{\rm flex}_{\rm int}$ ratio is very similar also for the same value of $R$ and $\theta=90^\circ$. However, one can observe that if the value of $R$ increases, the value of $\delta E^{\rm DBO}_{\rm int}$ decreases to zero faster than $\delta E^{\rm flex}_{\rm int}$. For instance, already for $R=12$~bohr, $\delta E^{\rm DBO}_{\rm int}$ amounts only 10\% of $\delta E^{\rm flex}_{\rm int}$, whereas for $14$~bohr this ratio drops below 4\%. This feature of the ratio between the $\delta E^{\rm DBO}_{\rm int}$ and $\delta E^{\rm flex}_{\rm int}$ corrections means that one can expect that adding $\delta E^{\rm DBO}_{\rm int}$ to $E^{\rm CCSD(T)}_{\rm int} + \delta E^{\rm FCI}_{\rm int} + \delta E^{\rm flex}_{\rm int}$ should not change significantly the calculated reaction rate coefficients, since in the major part of the range of the propagation of the wave function, the $\delta E^{\rm DBO}_{\rm int}$ correction is very small in comparison with other components of the interaction energy. Indeed, in Figure~\ref{fig_flex} one can see that the curve representing the rate coefficient calculated with the $E^{\rm CCSD(T)}_{\rm int} + \delta E^{\rm FCI}_{\rm int} + \delta E^{\rm flex}_{\rm int} + \delta E^{\rm DBO}_{\rm int}$ surface is very close to the curve representing the calculations with the $E^{\rm CCSD(T)}_{\rm int} + \delta E^{\rm FCI}_{\rm int} + \delta E^{\rm flex}_{\rm int}$ one. Concluding, the $\delta E^{\rm DBO}_{\rm int}$ correction has a tiny affect on the position and shape of the considered resonances.

\vspace{-0.15cm}
\begin{center}
\line(1,0){250}
\end{center}
\vspace{-1.5cm}

\begin{table}[b]
\caption{\label{tab1} Values of the interaction energy of He$^{\ast}$+H$_2$ obtained from the CCSD(T) calculations, $E_{\rm int}^{\rm CCSD(T)}$, and various corrections to this energy: $\delta E_{\rm int}^{\rm FCI}$ resulting from full configuration interaction, 0.4\%$E_{\rm int}^{\rm corr}$ equal to 0.4\% of the correlation part of the interaction energy, $\delta E_{\rm int}^{\rm flex}$ resulting from taking into account the H$_2$ nonrigidity effect, and $\delta E_{\rm int}^{\rm DBO}$ resulting from the use of the Born--Oppenheimer approximation. The values of $E_{\rm int}^{\rm CCSD(T)}$, $\delta E_{\rm int}^{\rm FCI}$, and 0.4\%$E_{\rm int}^{\rm corr}$ were calculated for purposes of Ref.~\cite{Klein:16}, while $\delta E_{\rm int}^{\rm flex}$ and $\delta E_{\rm int}^{\rm DBO}$ in this work. Two angular orientations, the linear one ($\theta=0^\circ$) and the T-shape one ($\theta=90^\circ$), and selected values of the intermolecular separation $R$ are represented. Energies are given in cm$^{-1}$.}
\begin{tabular}{c@{\quad}c@{\quad}c@{\quad}c@{\quad}c@{\quad}c@{\quad}c}\\
\hline \hline
$\theta$ ($\deg$) & $R$ (bohr) & $E_{\rm int}^{\rm CCSD(T)}$ & $\delta E_{\rm int}^{\rm FCI}$  & 0.4\%$E_{\rm int}^{\rm corr}$ & $\delta E_{\rm int}^{\rm flex}$ & $\delta E_{\rm int}^{\rm DBO}$  \\ 
\hline
 0 &\,   9.0 &\,\,\, 2.7023 &  -1.0109  &  -0.3276  &  -0.5909 &  \ 0.2675$^{*}$ \\[-0.25cm]
 0 &    10.0 &     -13.1925 &  -0.6402  &  -0.1848  &  -0.3694 &  0.0888         \\[-0.25cm]
 0 &    10.5 &     -14.3044 &  -0.4906  &  -0.1385  &  -0.2899 &  0.0562         \\[-0.25cm]
 0 &    11.0 &     -13.6141 &  -0.3741  &  -0.1041  &  -0.2274 &  0.0357         \\[-0.25cm]
 0 &    11.5 &     -12.0889 &  -0.2901  &  -0.0785  &  -0.1794 &  0.0227         \\[-0.25cm]
 0 &    12.0 &     -10.3777 &  -0.2226  &  -0.0594  &  -0.1419 &  0.0142         \\[-0.25cm]
 0 &    13.0 &\     -7.0985 &  -0.1358  &  -0.0348  &  -0.0854 &  0.0053         \\[-0.25cm]
 0 &    14.0 &\     -4.6985 &  -0.0846  &  -0.0210  &  -0.0522 &  0.0019         \\[-0.25cm]
 0 &    15.0 &\     -3.0959 &  -0.0542  &  -0.0131  &  -0.0320 &  0.0006         \\
\hline
 90 &\,   9.0 &\,\,\, 9.3624 & -1.4660  &  -0.2520  &  -0.2558 &  0.0721 \\[-0.25cm]
 90 &    10.0 &\,    -7.2224 & -0.8892  &  -0.1455  &  -0.1318 &  0.0281 \\[-0.25cm]
 90 &    10.5 &\,    -9.2548 & -0.6835  &  -0.1098  &  -0.0993 &  0.0184 \\[-0.25cm]
 90 &    11.0 &\,    -9.5490 & -0.5244  &  -0.0833  &  -0.0751 &  0.0117 \\[-0.25cm]
 90 &    11.5 &\,    -8.8666 & -0.4074  &  -0.0631  &  -0.0593 &  0.0075 \\[-0.25cm]
 90 &    12.0 &\,    -7.8076 & -0.3102  &  -0.0481  &  -0.0447 &  0.0048 \\[-0.25cm]
 90 &    13.0 &\,    -5.5574 & -0.1880  &  -0.0284  &  -0.0276 &  0.0018 \\[-0.25cm]
 90 &    14.0 &\,    -3.7612 & -0.1155  &  -0.0172  &  -0.0177 &  0.0007 \\[-0.25cm]
 90 &    15.0 &\,    -2.5106 & -0.0733  &  -0.0108  &  -0.0116 &  0.0002 \\
\hline  \hline
\multicolumn{7}{l}{$^{*}$extrapolated value}
\end{tabular}
\end{table}

\clearpage

\begin{table}[b] 
\vspace{0.2cm}
\begin{center}
\line(1,0){250}
\end{center}
\caption{\label{flexcomp} Second derivatives of the interaction energy components for the He$^*$+H$_2$ system for the atom--molecule distance of 10.5~bohr. To estimate the basis set limit, we used the d-aug-cc-pVQZ results for $E^{(1)}_{\rm elst}$, $E^{(1)}_{\rm exch}$, $E^{(2)}_{\rm ind}$, $E^{(2)}_{\rm exch-ind}$, and extrapolated values of $E^{(2)}_{\rm disp}$ and $E^{(2)}_{\rm exch-disp}$. For the induction and dispersion energies we used the TDHF model~\cite{Zuchowski_2003,Hapka_2012,Hapka13}, exchange and electrostatic interactions were obtained from the Hartree--Fock densities. The derivatives with respect to $r$ were calculated at $r=r_c$, in units cm$^{-1}$/(bohr)$^2$.}
\begin{center}
\begin{tabular}{l@{\quad}l@{\quad \qquad}r@{\qquad}r@{\qquad\qquad}r@{\qquad}rr} 
\hline\hline
 &   &\multicolumn{2}{l}{\quad  d-aug-cc-pVTZ}  &  \multicolumn{3}{l}{\negthinspace \negthinspace  \negthinspace \negthinspace  \negthinspace  estimated basis set limit}   \\  \hline
 \multicolumn{2}{l}{\quad\quad component}             &   $\theta=0^\circ$   & $\theta=90^\circ$   &   $\theta=0^\circ$   & $\theta=90^\circ$    \\ 
\hline
&  $\partial^2 E^{(1)}_{\rm elst}/\partial r^2$       &   -1.5729  & -0.0805  &  -1.3607   & -0.1609  \\ 
&  $\partial^2 E^{(1)}_{\rm exch}/\partial r^2$       &    7.3257  & -0.0086  &   7.2694   & -0.0313  \\ 
&  $\partial^2 E^{(2)}_{\rm ind}/\partial r^2$        &   -6.1842  &  0.0467  &  -6.1974   &  0.0326  \\
&  $\partial^2 E^{(2)}_{\rm exch-ind}/\partial r^2$   &    0.8964  & -1.0450  &   0.8999   & -1.0168  \\
&  $\partial^2 E^{(2)}_{\rm disp}/\partial r^2$       &  -23.1951  & -6.8707  & -23.8827   & -7.7385  \\            
&  $\partial^2 E^{(2)}_{\rm exch-disp}/\partial r^2$  &    2.1722  &  0.9165  &   2.3144   &  1.1485  \\
\hline
&  $\partial^2 V(R,\theta,r)/\partial r^2$            &  -20.5579  & -7.0416  & -20.9571   & -7.7666  \\ \hline\hline
\end{tabular}
\end{center}
\end{table}

Let us now discuss the uncertainty of the flexibility correction $\delta E_{\rm int}^{\rm flex}$. In Table~\ref{flexcomp} we gathered second derivatives of the components of the interaction energy for 10.5~bohr for T-shape and linear geometry. As one can see, the dispersion energy by far dominates the total flexibility correction. Given how good overall performance of SAPT was for the classically allowed region~\cite{Hapka_2012}, one can safely assume that the effect of higher order SAPT corrections will be marginal. There are two main uncertainties related to the dispersion derivative used in this paper: basis set incompleteness and the time-dependent Hartree--Fock (TDHF) approximation~\cite{Zuchowski_2003, Hapka_2012}. To address the first uncertainty, we performed test calculations using the d-aug-cc-pVQZ basis set for a few geometries around the minimum in T-shape and linear configurations. Using standard basis set extrapolation technique~\cite{Halkier_99} for the dispersion and the exchange--dispersion (note that only these components of the interaction energy should be extrapolated, since the others do not explicitly depend on excitations between subsystems) we found that for the linear geometry the second derivative is underestimated by about 2\%, while for T-shape by about 10\%. These numbers translate to about 6\% underestimation and small 3\% overestimation for the $V_0$ and $V_2$ parts of the $\delta E_{\rm int}^{\rm flex}$ correction, respectively. In absolute numbers these values are well below the uncertainty of basis set incompleteness for the $E_{\rm int}^{\rm CCSD(T)}$ part of the total interaction energy. To address the uncertainty of the TDHF method let us note that this model only very slightly overestimates the dispersion energy for the metastable helium dimer by about 3\%~\cite{Zuchowski_2003}. Similarly, the comparison of long-range isotropic $C_{60}$ coefficient obtained with TDHF~\cite{Hapka13} (112.8 $E_h a_0^6$) and the accurate value of Bishop and Pipin~\cite{BishopPipin_1993} (108.24 $E_h a_0^6$) suggests that, indeed, the dispersion energy can be slightly overestimated. Assuming that $\frac{\partial^2 E^{(2)}_{\rm disp}}{\partial r^2}\bigg|_{r=r_c}$ is proportional to overall performance of the dispersion, one can conclude that the inaccuracy due to the TDHF method is marginal and contributes to less than 0.01 cm$^{-1}$ in the minimum range.

\vspace{0.5cm}